\documentclass[twocolumn, 10pt, amsmath, amssymb, showpacs]{revtex4}
\usepackage{epsfig, amssymb, times, graphicx, epstopdf}

\newcommand{\finpr}{\hfill $\square$ \vspace{2mm}}

\newcommand{\ket}[1]{|#1\rangle}

\def\be{\begin{eqnarray}}
\def\ee{\end{eqnarray}}
\def\bee{\begin{eqnarray*}}
\def\eee{\end{eqnarray*}}
\newtheorem{thm}{Theorem}

\newtheorem{defn}{Definition}

\def\ket{\rangle}

\newcommand{\ketbra}[2]{\ensuremath{| #1 \rangle \langle #2 |}}

\begin{document}

\title{Graph states as ground states of many-body spin-$1/2$ Hamiltonians}

\author{M. Van den Nest$^1$, K. Luttmer$^1$, W. D\"ur$^{1,2}$, H. J. Briegel$^{1,2}$}

\affiliation{$^1$ Institut f\"ur Quantenoptik und Quanteninformation der \"Osterreichischen Akademie der Wissenschaften, Innsbruck, Austria\\
$^2$ Institut f{\"u}r Theoretische Physik, Universit{\"a}t Innsbruck,
Technikerstra{\ss}e 25, A-6020 Innsbruck, Austria
}
\date{\today}

\date{\today}
\def\makeheadbox{}

\begin{abstract}
We consider the problem whether  graph states can be ground
states of local interaction Hamiltonians. For Hamiltonians
acting on $n$ qubits that involve at most two-body
interactions, we show that no $n$-qubit graph state can be
the exact, non-degenerate ground state. We determine for
any graph state the minimal $d$ such that it is the
non-degenerate ground state of a $d$-body interaction
Hamiltonian, while we show for $d'$-body Hamiltonians $H$
with $d'<d$ that the resulting ground state can only be
close to the graph state at the cost of $H$ having a small
energy gap relative to the total energy.  When allowing for
ancilla particles, we show how to utilize a gadget
construction introduced in the context of the $k$-local
Hamiltonian problem, to obtain $n$-qubit graph states as
non-degenerate (quasi-)ground states of a two-body
Hamiltonian acting on $n'>n$ spins.
\end{abstract}
\maketitle

\section{Introduction}
Graph states form a large family of  multi-particle quantum
states that are associated with mathematical graphs. More
precisely, to any $n$-vertex graph $G$ an $n$-qubit graph
state $|G\rangle$ is associated. Graph states play an
important role in several applications in quantum
information theory and quantum computation \cite{He06}.
Most prominently, the graph states that correspond to a two
dimensional square lattice, also known as the 2D cluster
states \cite{Br01}, are known to be universal resources for
measurement based quantum computation \cite{Ra01, Ra03}.
Other graph states serve as algorithmic specific resources
for measurement based quantum computation, are codewords of
error correcting codes known as Calderbank-Shor-Steane
codes \cite{Ca96, St96}, or are used in multiparty communication schemes
\cite{Hi99, Ch04, Du05}. Graph states are specific instances of stabilizer
states, which allows for an efficient description and
manipulation of the states by making use of the stabilizer
formalism \cite{Go97}. This also makes them attractive as
test-bed states to investigate the complex structure of
multi-partite entangled quantum systems \cite{He06}. We
refer to Ref. \cite{He06} for an extensive review about
graph states and their applications.

While it is known how to efficiently  prepare $n$-qubit
graph states in highly controlled quantum systems using
only $O(n^2)$ phase gates, the question whether certain
graph states can occur naturally as non-degenerate ground
states of certain, physically reasonable, local interaction
Hamiltonians remains unanswered so far. This question is of
direct practical relevance, as if this would be the case,
graph states could simply be prepared by cooling a system
governed by such a local interaction Hamiltonian, without
further need to access or manipulate individual qubits in a
controlled way. Moreover, measurement based quantum
computation could then be realized by simply cooling and
measuring.

The first results in this context have recently been
obtained by Nielsen in Ref. \cite{Ni05}, where classes of
graph states which \emph{cannot} possibly occur as ground
states of two-body Hamiltonians have been constructed.
Moreover, numerical evidence was reported which raised the
conjecture that in fact no $n$-qubit graph state can occur
as the non-degenerate ground state of an $n$-qubit two-body
spin-$1/2$ Hamiltonian---and this conjecture is believed to
hold by several researchers in the field. Nonetheless, a
definitive answer to the question whether there exists
graph states which are non-degenerate ground states of
two-body Hamiltonians, is to date missing. In this article
we provide a proof of this conjecture for arbitrary
$n$-qubit graph states and two-body interaction
Hamiltonians acting on $n$ qubits.

In fact, we provide some more general results, where we
consider $n$-qubit graph states $|G\rangle$ that are ground
states of $d$-body Hamiltonians $H$ acting on $n$ qubit
systems, where $d$ is possibly larger than two. We find the
following:
\begin{itemize}
\item[(i)] For every graph state $|G\rangle$, we determine
the minimal $d$ such $|G\rangle$ is a, possibly degenerate,
ground state of a $d$-body Hamiltonian $H$, and show that
$d$ is equal to the minimal weight of stabilizer group of
$|G\rangle$. \item[(ii)] For every graph state $|G\rangle$,
we determine the minimal $d$ such that $|G\rangle$ is the
{\em non-degenerate} ground state of a $d$-body Hamiltonian
$H$, where we show that $d$ is again related to the
stabilizer via a quantity which we denote by
$\eta(|G\rangle)$. In addition, we find that
$\eta(|G\rangle)$ cannot be smaller than 3 for all graph
states, i.e., \emph{no graph state can be the
non-degenerate ground state of a two-body hamiltonian.}
\item[(iii)]  For $d' < \eta(|G\rangle)$, we show that the
ground state of any $d'$-body hamiltonian $H$ can only be
$\epsilon$-close to $|G\rangle$ at the cost of $H$ having
an energy gap (relative to the total energy) that is proportional to $\epsilon$.
\end{itemize}

If we allow for ancilla particles,  i.e., consider systems
of $n' > n$ spins where we are interested only in the state
of a subset of $n$ qubits, we find that
\begin{itemize}
\item[(iv)] Any graph  state $|G\rangle$ of $n$ qubits can
be the non-degenerate (quasi-)ground state of a two-body
Hamiltonian that acts on $n' > n$ qubits.
\end{itemize}
The latter result follows from a so-called \emph{gadget
construction} introduced in Ref. \cite{Ke04, Ol05} in the context of the
$k$-local Hamiltonian problem. Here, the ancilla particles
act as mediating particles to generate an effective
many-body Hamiltonian on the $n$ system particles, using
only two-body interactions. Although this leads in
principle to a way to obtain graph states as non-degenerate
ground states of two-body interaction Hamiltonians, a high
degree of control is required in the interaction
Hamiltonians as the parameters describing the interaction
need to be precisely adjusted. We remark that this result
is closely related to a recent finding of Bartlett and
Rudolph \cite{Ba06}, who show how to obtain an {\em
encoded} graph state corresponding to a universal resource
for measurement based quantum computation (a 2D cluster
state or a state corresponding to a honeycomb lattice
\cite{Va06}) using a construction based on projected
entangled pairs.

The paper is organized as follows. In section
\ref{sect_defs}, we provide definitions and settle
notation. In section \ref{sect_deg}, we consider degenerate
ground states and provide a simple argument to determine
the minimal $d$ such that $|G\rangle$ can be the exact
ground state of a $d$-body Hamiltonian. In section
\ref{sect_nondeg} we treat the non-degenerate case, and
compute, for every graph state, the minimal $d$ such that
this graph state is the non-degenerate ground state of a
$d$-body Hamiltonian; this quantity is denoted by
$\eta(|G\rangle)$. In section \ref{sect_approx} we consider
approximate ground states, i.e. approximations of
$|G\rangle$ using $d'$-body Hamiltonians where $d'$ is
strictly smaller than $\eta(|G\rangle)$. In section
\ref{sect_ancilla} we consider the case of ancilla
particles, and show how to obtain any $n$-qubit graph state
as a non-degenerate (quasi-)ground state of a two-body
Hamiltonian on $n'>n$ qubits. We demonstrate the gadget
construction for the one-dimensional cluster state and the
honeycomb lattice. Finally, we conclude in section
\ref{sect_concl}.

\section{Definitions and notations}\label{sect_defs}

The Pauli  spin matrices are denoted by $\sigma_x,
\sigma_y, \sigma_z$, and $\sigma_0$ denotes the $2\times 2$
identity operator. A Pauli operator on $n$ qubits is an
operator of the form $\sigma=
\sigma_{i_1}\otimes\dots\otimes\sigma_{i_n},$ where $i_1,
\dots, i_n\in\{0,x,y,z\}$. The \emph{weight} wt$(\sigma)$
of $\sigma$ is the number of qubit systems on which this
operator acts nontrivially. A $d$-body spin-$1/2$
Hamiltonian on $n$ qubits is a Hermitian operator of the
form $H = \sum_{\sigma} h_{\sigma} \sigma,$ where the sum
runs over all $n$-qubit Pauli operators $\sigma$ with
wt$(\sigma)\leq d$, and where the $h_{\sigma}$ are real
coefficients.

Let $G=(V, E)$ be a graph with vertex set $V=\{1, \dots,
n\}$ and edge set $E$. The graph state $|G\rangle$ is
defined to be the simultaneous fixed point of the $n$
operators \be\label{graphstate}\sigma^{(a)}\prod_{b\in
N(a)}\sigma^{(b)},\ee for every $a\in V$, where
$N(a)\subseteq V$ denotes the set of vertices $b$ connected
to $a$ by an edge. The superscripts denote on which system
a Pauli operator acts. The stabilizer ${\cal S}$ of
$|G\rangle$ is the group of all operators of the form
$g=\pm\sigma$, where $\sigma$ is a Pauli operator on $n$
qubits, satisfying $g|G\rangle=|G\rangle$.  Recall that
$|{\cal S}|=2^n$ and that ${\cal S}$ is an Abelian group,
i.e., $[g, g']=0$ for every $g, g'\in{\cal S}$. We will
frequently use the expansion \be\label{exp}|G\rangle\langle
G| = \frac{1}{2^n}\sum_{g\in {\cal S}} g.\ee We refer to
Ref. \cite{He06} for extensive material regarding graph
states and the stabilizer formalism.

To avoid technical issues arising in trivial cases, we will
only consider fully entangled graph states (corresponding
to connected graphs) on $n\geq 3$ qubits.

\section{Degenerate ground states}\label{sect_deg}

In this section we investigate under which conditions a
graph state is the ground state of a $d$-body Hamiltonian.
At this point we will not yet require that this ground
state is non-degenerate---this case will be  considered
below---which is a considerable simplification, and we will
see that elementary arguments suffice to gain total insight
in this matter. In sections \ref{sect_deg_th} and
\ref{sect_deg_delta} these insights are obtained, and
examples are given in section \ref{sect_deg_ex}.

\subsection{General results}\label{sect_deg_th}

In order to exclude trivial cases, we will only consider
Hamiltonians which are both nonzero and not a multiple of
the identity; such Hamiltonians will be called
\emph{nontrivial}. We will need the following definition.

\begin{defn}
Letting $|G\rangle$ be an $n$-qubit graph state with
stabilizer ${\cal S}$, define
\be\delta(|G\rangle):=\min_{g\in{\cal S}\setminus\{I\}}
\mbox{wt}(g), \ee i.e., $\delta(|G\rangle)$ is defined to
be the minimal weight of ${\cal S}$.
\end{defn}
One  immediately finds that the nontrivial
$\delta(|G\rangle)$-body Hamiltonian \be H:=
-\sum_{g\in{\cal S},\
\mbox{\scriptsize{wt}}(g)=\delta(|G\rangle)} g\ee has the
state $|G\rangle$ as a ground state. Furthermore, let
$d<\delta(|G\rangle)$ and suppose that $H'$ is a nontrivial
$d$-body Hamiltonian having $|G\rangle$ as a ground state.
We prove that this leads to a contradiction. Defining the
nontrivial Hamiltonian \be H'':= H'-2^{-n}\mbox{
Tr}(H')I,\ee it follows that Tr$(H'')=0$ and that
$|G\rangle$ is also a ground state of $H''$. Letting $E_0$
be the ground state energy of $H''$, it follows from
(\ref{exp}) that \be\label{delta} E_0= \langle
G|H''|G\rangle = \frac{1}{2^n}\sum_{g\in{\cal S}}\mbox{
Tr}(gH'').\ee As $d<\delta(|G\rangle)$ and $H''$ is a
$d$-body Hamiltonian, one has $\mbox{ Tr}(gH'')=0$ for
every $I\neq g\in{\cal S}$. Furthermore, one has \be\mbox{
Tr}(H''\cdot I) = \mbox{ Tr}(H'') =0\ee by construction of
$H''$, and therefore $\mbox{ Tr}(gH'')=0$ for \emph{every}
$g\in{\cal S}$. Together with (\ref{delta}), this shows
that $E_0=0$. As $H''$ has zero trace and $E_0=0$ is the
smallest eigenvalue, this implies that $H''=0$, yielding a
contradiction, since $H'$ was assumed to be nontrivial.
This shows that $|G\rangle$ cannot be the ground state of
$H'$.

We have proven the following result:

\begin{thm}\label{thm_deg}
Let $|G\rangle$ be a graph state on $n$ qubits and let
$\delta(|G\rangle)$ be defined as above. Then
\begin{itemize}
\item[(i)] there exists a  nontrivial
$\delta(|G\rangle)$-body Hamiltonian on $n$ qubits having
$|G\rangle$ as a---possibly degenerate---ground state;
\item[(ii)] any nontrivial $n$-qubit Hamiltonian having
$|G\rangle$ as a ground state must involve at least
$\delta(|G\rangle)$-body interactions.\end{itemize}
\end{thm}
Thus, $\delta(|G\rangle)$ is the optimal $d$ such that the
graph state $|G\rangle$ is the ground state of a nontrivial
$d$-body Hamiltonian. Note that some graph states can occur
as degenerate ground states of two-body Hamiltonians,
namely  whenever the stabilizer of $|G\rangle$ contains
elements of weight 2. This e.g. occurs when the graph has
one or more vertices of degree 1 (see (\ref{graphstate})).
We refer to section \ref{sect_deg_ex} for examples.

In the next section we focus on some techniques to compute
$\delta(|G\rangle)$.

\subsection{Computing $\delta(|G\rangle)$}\label{sect_deg_delta}

We will show that $\delta(|G\rangle)$ can be computed
directly from the graph $G$ in two distinct ways, one
method being algebraic and the other graphical.

In the graphical approach, one considers a graph
transformation rule called \emph{local complementation},
defined as follows. Let $a\in V$ be a vertex of $G$ and let
$N(a)\subseteq V$ denote the neighborhood of this vertex.
The local complement $G*a$ of the graph $G$ at the vertex
$a$ is defined to be the graph obtained by complementing
(i.e., replacing edges with non-edges and vice-versa) the
subgraph of $G$ induced on the subset $N(a)$ of vertices,
and leaving the rest of the graph unchanged. Local
complementations of graphs correspond to local (Clifford)
operations on the corresponding graph states \cite{Va04}.
Moreover, it was proven in Ref. \cite{Mh04} that
$\delta(|G\rangle)-1$ is equal to the minimal vertex degree
of any graph which can be obtained from $G$ by applying
local complementations. This immediately yields a graphical
method to calculate $\delta(|G\rangle)$, as one simply has
to draw all graphs which can be obtained from $G$ by local
complementations and determine the smallest vertex degree
in this class of graphs. We refer to Refs. \cite{He06,
Va04} for more details on local complementation of graphs
and local operations on graph states.

In the  algebraic approach, one considers the adjacency
matrix $\Gamma$ of $G$ and defines, for every subset
$A\subseteq V$ of the vertex set of $G$, the $|A|\times
(n-|A|)$ matrix \be\Gamma\langle A\rangle: = \left(
\Gamma_{ab}\right)_{a\in A, b\in V\setminus A}.\ee We can
then formulate the following result.

\begin{thm}\label{thm_delta}
Let $G=(V, E)$  be a graph with adjacency matrix $\Gamma$.
Then $\delta(|G\rangle)$ is equal to the smallest possible
cardinality $|A|$ of a subset $A\subseteq V$ such that
$|A|> \mbox{rank}_{2}\ \Gamma\langle A\rangle$ (where
'rank$_2\ X$' denotes the rank of a matrix $X$ over the
finite field GF(2).)
\end{thm}
{\it Proof: } the proof uses standard  graph state
techniques, see Ref. \cite{He06}. First one shows that $ 2^{|A|-\mbox{
rank}_2\Gamma\langle A\rangle}$ is equal to the number of
elements $g=\pm\sigma_{i_1}\otimes\dots\otimes\sigma_{i_n}$
in the stabilizer ${\cal S}$ of $|G\rangle$ satisfying
$i_{\alpha}=0$ for every $\alpha\in V\setminus A$.
Therefore, if $|A|=\mbox{rank}_{2}\ \Gamma\langle A\rangle$
for every subset $A$ with cardinality $|A|=d$, then ${\cal
S}$ does not contain elements of weight $d$ or less. The
result then immediately follows. \finpr

Note that the  above result yields a polynomial time
algorithm to test, for a given constant $k$ (independent of $n$), whether $\delta(|G\rangle)$ is
smaller than $k$. In order to do so, one needs to calculate
the rank over GF(2) of all matrices $\Gamma\langle
A\rangle$ with $|A|\leq k$; since the number of such
matrices is $\sum_{i=0}^k \binom{n}{i}$ (which is
polynomial in $n$) and the rank of a matrix can be
calculated in polynomial time in the dimensions of the
matrix, one obtains a polynomial algorithm to verify
whether $\delta(|G\rangle)$ is smaller than $k$. However, a direct computation of $\delta(|G\rangle)$ is likely to be a hard problem.

\subsection{Examples}\label{sect_deg_ex}
Here we give some examples of the theoretical results
obtained regarding  graph states as non-degenerate ground
states.

Following theorem \ref{thm_delta}, one finds that a graph
state is the---degenerate---ground state of a two-body
Hamiltonian if and only if its stabilizer contains elements
of weight 2. For example, the GHZ state $(|0\rangle^n +
|1\rangle^n)/\sqrt{2}$ (which is locally equivalent to the
graph state defined by the fully connected graph) is the
degenerate ground state of the Ising Hamiltonian with zero
magnetic field: \be H = -\sum_{i=1}^{n-1}
\sigma_z^{(i)}\sigma_z^{(i+1)}.\ee One can easily verify
that this ground state is 2-fold degenerate.

The linear cluster  $|L_n\rangle$ state with open boundary
conditions, where $L_n$ is the linear chain on $n$
vertices, also is the degenerate ground state of a two-body
Hamiltonian. This immediately follows by considering the
stabilizer generators associated to the 2 boundary vertices
of the  graph $L_n$, which both have degree 1, therefore
$\delta(|L_n\rangle)=2$ by using the graphical approach to
determine $\delta(|G\rangle)$, as explained in section
\ref{sect_deg_th}. The state $|L_n\rangle$ is the ground
state of the Hamiltonian \be -\sigma_x^{(1)}\sigma_z^{(2)}
-\sigma_z^{(n-1)}\sigma_x^{(1)}.\ee However, the degeneracy
of the ground state energy is very large, namely $2^{n-2}$.
(Moreover, an argument similar to the proof of theorem
\ref{thm_nondeg} shows that any two-body Hamiltonian having
$|L_n\rangle$ as a ground state must exhibit at least this
degeneracy).

Consider the 1D cluster state $|C_n\rangle$ with periodic
boundary conditions, which is a graph state where the
underlying graph is a cycle graph $C_n$. The adjacency
matrix of $C_n$ is given by \be\Gamma =
\left[\begin{array}{cccccc}
\cdot&1&\cdot&\cdot&\cdot&1\\
1&\cdot&1&\cdot&\cdot&\cdot\\
\cdot&1&\cdot&1&\cdot&\cdot\\
\cdot&\cdot&1&\cdot&1&\cdot\\
\cdot&\cdot&\cdot&1&\cdot&1\\
1&\cdot&\cdot&\cdot&1&\cdot\end{array} \right],\ee were we
give the example for $n=6$. The periodic boundary
conditions are chosen such as to eliminate boundary
effects. Using theorem \ref{thm_delta}, we easily find that
$\delta(|C_n\rangle)=3$. Indeed, $\Gamma\langle
\{a\}\rangle$ has rank 1 for every $a\in V$, and
$\Gamma\langle \{a, b\}\rangle$ has rank 2 for every
2-element subset $\{a, b\}$ of $V$. Moreover, the rank of
\be \Gamma\langle \{1, 2, 3\}\rangle =
\left[\begin{array}{ccc}
\cdot&\cdot&1\\
\cdot&\cdot&\cdot\\
1&\cdot&\cdot\end{array} \right]\ee is equal to 2, and
therefore $\delta(G)=|\{1,2,3\}|=3.$

The 2D  cluster state $|C_{k\times k}\rangle$ has
$\delta(|C_{k\times k}\rangle)=3$ (open boundary
conditions) or $\delta(|C_{k\times k}\rangle)=5$ (periodic
boundary conditions), which can be verified by applying
theorem \ref{thm_delta}.

\section{Non-degenerate ground states}\label{sect_nondeg}

While in the previous section we have found that graph
states can be degenerate ground states of two-body
Hamiltonians, next we show that they can never be
non-degenerate ground states of such Hamiltonians. As in
the previous section, first we present the general results
in section \ref{sect_nondeg_th}, after which we give
examples in section \ref{sect_nondeg_ex}.

\subsection{General results}\label{sect_nondeg_th}

In order to investigate non-degenerate ground states and
their relation to graph states, we will need the following
definition.

\begin{defn}
Let $|G\rangle$ be a graph state on $n$ qubits with
stabilizer ${\cal S}$. Let ${\cal S}_d$ be the subgroup of
${\cal S}$ generated by all elements of weight at most $d$
\cite{foot}.
 Then $\eta(|G\rangle)$ is defined to be the minimal $d$ such that
${\cal S}_d= {\cal S}$.
\end{defn}
We can now state the second main result of this article.
\begin{thm}\label{thm_nondeg}

Let $|G\rangle$ be a graph state on $n\geq 3$ qubits, and
let $\eta(|G\rangle)$ be defined as above. Then
\begin{itemize}
\item[(i)] there exists an $\eta(|G\rangle)$-body
Hamiltonian on $n$ qubits having $|G\rangle$ as a
non-degenerate ground state; \item[(ii)] any $n$-qubit
Hamiltonian having $|G\rangle$ as a non-degenerate ground
state must involve at least $\eta(|G\rangle)$-body
interactions. \item[(iii)] $\eta(|G\rangle)\geq 3$, i.e.,
no $n$-qubit graph state is the non-degenerate ground state
of a two-body Hamiltonian on $n$ qubits.
\end{itemize}
\end{thm}
{\it Proof: } first we show that $|G\rangle$ is the
non-degenerate ground state of an $\eta(|G\rangle)$-body
Hamiltonian. To see this, note that by definition of
$\eta(|G\rangle)$ there exists a set of generators $\{g_1,
\dots, g_n\}$ of ${\cal S}$ such that wt$(g_i)\leq
\eta(|G\rangle)$ for every $i=1, \dots, n$. Therefore, the
Hamiltonian $H := -\sum_{i=1}^n g_i$ involves at most
$\eta(|G\rangle)-$body interactions. Moreover, $H$ has the
state $|G\rangle$ as a non-degenerate ground state. To see
this, note that the operators $g_i$ mutually commute, and
that they have eigenvalues $\pm 1$. Therefore, the smallest
eigenvalue of $H$ is equal to $-n$. As $H|G\rangle =
-n|G\rangle$ trivially, this shows that $|G\rangle$ is a
ground state of $H$. Furthermore, this ground state is
non-degenerate, as any state $|\psi\rangle$ satisfying
$H|\psi\rangle = -n|\psi\rangle$ must also satisfy
$g_i|\psi\rangle = |\psi\rangle$ for every $i=1, \dots, n$,
and therefore $|\psi\rangle$ must be equal to $|G\rangle$
up to a global phase. This shows that $H$ has $|G\rangle$
as non-degenerate ground state.

Conversely, we prove that any Hamiltonian having
$|G\rangle$ as a non-degenerate ground state must involve
at least $\eta(|G\rangle)-$body terms. To see this, suppose
that $H'$ is a $d$-body Hamiltonian, with $d< \eta(|G\rangle)$,
having $|G\rangle$ as a ground state. Let $\{g_1, \dots,
g_s\}$ be a set of independent \cite{foota} generators
 of ${\cal S}_d$, where $s=\log_2 |{\cal S}_d|$.  As $d<\eta(|G\rangle)$ it
follows that ${\cal S}_d$ cannot be equal to ${\cal S}$.
Hence there exists a non-empty independent set of elements
$\{g_{s+1}, \dots, g_n\}\subseteq {\cal S}$, where
wt$(g_i)\geq d+1$ for every $i=s+1, \dots, n$, such that
$\{g_1, \dots, g_n\}$ is an (independent) generating set of
${\cal S}$. We then define $|G; \gamma\rangle$ to be the
stabilizer state with stabilizer ${\cal S}_{\gamma}$
generated by the set \be \{g_1, \dots, g_s,
(-1)^{\gamma_{s+1}} g_{s+1}, \dots, (-1)^{\gamma_n}
g_{n}\},\ee for every $ \gamma:=(\gamma_{s+1}, \dots,
\gamma_n)\in \{0,1\}^{n-s}$.

We now claim that
\be\label{3cases}\langle G|\tau| G\rangle = \langle
G;\gamma|\tau| G;\gamma\rangle\ee for every Pauli operator
$\tau$ of weight at most $d$. This property can be shown by
considering (\ref{exp}) and a similar expansion for
$|G;\gamma\rangle$, and making a distinction between the
following cases.
\begin{itemize}
\item[(a)] $\alpha\tau\in{\cal S}$ for some $\alpha=\pm 1$;
\item[(b)] both $\tau$ and $-\tau$ do not belong to ${\cal S}$;
\end{itemize}
First, if (a) $\alpha\tau\in{\cal S}$ for some
$\alpha=\pm 1$ then by construction $\alpha\tau\in{\cal
S}_{\gamma}$, and therefore \be\langle G|\tau| G\rangle =
\langle G;\gamma|\tau| G;\gamma\rangle =\alpha\ee
Second, if (b)
both $\tau$ and $-\tau$ do not belong to ${\cal S}$, then
none of these two operators can belong to the stabilizer
${\cal S}_{\gamma}$: for, suppose that $\alpha \tau\in{\cal
S}_{\gamma}$ for some $\alpha=\pm 1$; then $\alpha\tau$ can
be written in a unique way as a product
\be\alpha\tau=\prod_{i=1}^s g_i^{a_i}
\prod_{j=s+1}^n((-1)^{\gamma_{j}} g_{j})^{a_{j}}\ee for
some $(a_1\dots, a_n)\in\{0,1\}^n$. But then clearly either
$\tau$ or $-\tau$ is equal to $g_1^{a_1}\dots
g_n^{a_n}\in{\cal S}$, which yields a contradiction. We can
now conclude, as both $\tau$ and $-\tau$ belong to neither
${\cal S}$ nor ${\cal S}_{\gamma}$, that \be\langle G|\tau|
G\rangle = \langle G;\gamma|\tau| G;\gamma\rangle=0.\ee This
proves property (\ref{3cases}). Using this identity, we
find that $\langle G|H'| G\rangle=\langle G;\gamma|H'|
G;\gamma\rangle$ for every $\gamma$. As $\langle G|H'|
G\rangle$ is equal to the ground state energy $E_0$ of
$H'$, this shows that every state $|G;\gamma\rangle$ is an
eigenstate of $H'$ with eigenvalue $E_0$. Hence, the ground
state of $H'$ is degenerate.

Finally, it we prove that ${\cal S}_2$ cannot be
equal to ${\cal S}$ for any (fully entangled) graph state on $n\geq 3$
qubits, which implies that $\eta(|G\rangle)\geq 3$.

Note that a fully entangled graph state cannot have stabilizer elements of weight 1. Suppose that $g_1, \dots, g_n$ are $n$ Pauli operators of weight 2 which generate ${\cal S}$. We will show that this leads to a contradiction. For two arbitrary such operators $g_i$ and $g_j$, we distinguish between three possibile cases: (i) $g_i$ and $g_j$ act nontrivially on disjoint pairs of qubits, (ii) $g_i$ and $g_j$ act nontrivially on the same pair of qubits, and (iii) $g_i$ acts nontrivially on qubits $a$ and $b$, and $g_j$ acts on qubits $a$ and $c$, for some $a, b, c\in\{1, \dots, n\}$. If (ii) is the case, one finds that the operators $g_i$, $g_j$ and $g_ig_j$, which act on the same pair of qubits, belong to the stabilizer. Due to the commutativity of these operators, one finds that up to a local unitary operation, one has $g_i = X\otimes X\otimes I_{n-2}$, $g_j = Z\otimes Z\otimes I_{n-2}$, and $g_ig_j = -Y\otimes Y\otimes I_{n-2}$, where $I_k$ is the identity on $k$ qubits.  As \be \{I_2, X\otimes X, Z\otimes Z, -Y\otimes Y\}\ee is the complete stabilizer of a two-qubit (Bell) state $|\psi_B\rangle$, it then follows that $|G\rangle$ can be written as a tensor product $|G\rangle = |\psi_B\rangle\otimes|\tilde\psi\rangle$ for some $|\tilde\psi\rangle$. This yields a contradiction, as we have assumed that $|G\rangle$ is fully entangled. Thus, only cases (i) and (iii) occur. It is then easy to show that any set of independent Pauli operators which satisfies these conditions, can  contain at most $n-1$ elements. However, $n$ elements are required to obtain a full generating set of a stabilizer. This shows that no fully entangled graph state can be generated by weight two elements.
This proves the result. \finpr

A few remarks are in order. First, the above result proves
that graph states cannot be non-degenerate ground states
of two-body Hamiltonians. This settles the question raised by Nielsen in Ref. \cite{Ni05}.

Second, it follows from the proof of theorem
\ref{thm_nondeg} that, given any $d<\eta(|G\rangle)$ and
any Hamiltonian $H'$ having $|G\rangle$ as a (necessarily
degenerate) ground state, then the degeneracy is at least
equal to \be\label{degeneracy} 2^{n-s}= 2^n |{\cal
S}_d|^{-1},\ee since we have proven that the $2^{n-s}$
orthogonal states $\{|G;\gamma\rangle\}_{\gamma}$ are all
ground states of $H'$. Also, note that the d-body
Hamiltonian \be H'':= -\sum_{g\in{\cal S},
\mbox{\scriptsize{ wt}(g)}\leq d} g\ee has $|G\rangle$ as a
degenerate ground state, where the degeneracy is exactly
equal to (\ref{degeneracy}).

Finally, we note that computing $\eta(|G\rangle)$ for an
arbitrary graph state is likely to be hard. A brute force
approach would be the following: enumerate all generating
sets $S:=\{g_1, \dots, g_n\}$ of the stabilizer ${\cal S}$
of $|G\rangle$, and determine wt$(S)$, which is defined to
be the maximal weight of an element in $S$. Then the
minimal value of wt$(S)$, when $S$ ranges over of all
generating sets, is then equal to $\eta(|G\rangle)$.
Clearly this approach is non-polynomial, as the stabilizer
of a graph state on $n$ qubits has ${\cal O}(2^{n^2})$
generating sets. Nevertheless, in the next section we will
encounter some interesting examples of graph states where
$\eta(|G\rangle)$ can be computed quickly.

\subsection{Examples}\label{sect_nondeg_ex}

In this section we consider some examples of the
calculation of $\eta(|G\rangle)$. Note that one always has
$\eta(|G\rangle)\geq \delta(|G\rangle)$.

Consider the linear cluster state with periodic boundary
conditions. The stabilizer of the state $|C_n\rangle$ is
generated by the elements \be
\sigma_z^{(a-1)}\sigma_x^{(a)}\sigma_z^{(a+1)}\ee ($a=1,
\dots, n.$) This implies that $|C_n\rangle$ is the
nondegenerate ground state of the 3-body Hamiltonian
\be\label{H_C} -\sum_{a=1}^n\
\sigma_z^{(a-1)}\sigma_x^{(a)}\sigma_z^{(a+1)}.\ee  Theorem
\ref{thm_nondeg}(iii) then shows that $H$ is optimal in the
sense that no 2-body Hamitonian exists having $|C_n\rangle$
as a ground state. Thus, we have proven that
$\eta(|C_n\rangle)=3$.

For the linear cluster state with open boundary conditions
$|L_n\rangle$ one also finds that
$\eta(|L_n\rangle)=3$, as $|L_n\rangle$ is the
non-degenerate ground state of a three-body Hamiltonian
analogous to  (\ref{H_C}).

As for the 2D cluster states, one finds that
$\eta(|C_{k\times k}\rangle) = 5$, showing that at
least 5-body interactions are needed to have these states
as non-degenerate ground states. We note that Nielsen
already proved in Ref. \cite{Ni05} that the 2D cluster
states on $n$ qubits cannot be ground states of 2-body
Hamiltonians on $n$ qubits.

\section{Approximate ground states}\label{sect_approx}

Having determined that at least $\eta(|G\rangle)$-body
interactions are needed to obtain a graph state $|G\rangle$
as an \emph{exact} non-degenerate ground state, next we
investigate whether it is possible to obtain non-degenerate
ground states \emph{close to} graph states when only
$d-$body Hamiltonians are considered with
$d<\eta(|G\rangle)$. We will show that this only possible
if the Hamiltonians have small energy gaps between the
ground state and the first excited level. In order to
obtain this result, we prove a technical result which
relates the fidelity between a graph state and the ground
state of a Hamiltonian, and the spectrum of this
Hamiltonian.

\begin{thm}\label{thm_techn}
Let $d$ be a positive integer. Let $|G\rangle$ be an
$n$-qubit graph state with stabilizer ${\cal S}$, let
${\cal S}_d\subseteq{\cal S}$ be defined as above and
denote $r:=2^n|{\cal S}_d|^{-1}$. Let $H$ be a $d$-body,
$n$-qubit Hamiltonian with ground state $|\psi\rangle$, and
let ${\bf E}=(E_0, E_1, \dots, E_{2^n-1})$ be the energies
of $H$ in ascending order. Then \be \frac{1}{\sqrt{2} \|
{\bf E}\|}\left( \frac{E_0+\dots + E_{r-1}}{r} -
E_0\right)\leq (1-|\langle
G|\psi\rangle|^2)^{1/2}.\nonumber\ee
\end{thm}
{\it Proof: } consider the Hermitian operator \be\rho_d :=
\frac{1}{2^n}\sum_{g\in{\cal S}_d} g.\ee Note that this
operator satisfies \be(\rho_d)^2 &=& \frac{1}{2^{2n}}
\sum_{g\in{\cal S}_d} g \sum_{h\in{\cal S}_d}h\nonumber\\
&=& \frac{1}{2^{2n}}\sum_{g\in{\cal S}_d} \sum_{h\in{\cal
S}_d}h\nonumber\\ &=& \frac{|{\cal S}_d|}{2^n} \rho_d =
r^{-1}\rho_d.\ee The second equality holds since ${\cal
S}_d$ is a group. It follows that $(r\rho_d)^2=r\rho_d$,
showing that $r\rho_d$ a projection operator. Thus, all
nonzero eigenvalues of this operator are equal to 1,
implying that the trace of $r\rho_d$ is equal to the rank
of this matrix. As the trace of $\rho_d$ is equal to 1, it
follows that the rank of $\rho_d$ (and $r\rho_d$) is equal
to $r$.

We now use Ky Fan's maximum principle, which states the
following: letting $H$ be any Hermitian operator, the
minimum value of Tr$(PH)$, when the minimization is taken
over all projection operators $P$ of rank $r$, is equal to
the sum of the $r$ smallest eigenvalues $E_0,E_1, \dots,
E_{r-1}$ of $H$. As $r\rho_d$ is a rank $r$ projector, this
shows that \be E_0+ E_1+\dots + E_{r-1}\leq r\mbox{
Tr}(\rho H).\ee Note that $\mbox{Tr}(\tau\rho_d) = \langle
G|\tau|G\rangle$ for every Pauli operator $\tau$ of weight
at most $d$, and we therefore have $\mbox{Tr}(\rho_d H) =
\langle G|H|G\rangle$, showing that \be\label{Ky}
\frac{1}{r}(E_0+ E_1+\dots + E_{r-1})\leq \langle
G|H|G\rangle .\ee We now determine an upper bound to
$\langle G|H|G\rangle$ in terms of the fidelity
$F:=|\langle G|\psi\rangle|$. By the Cauchy-Schwarz
inequality the trace $\mbox{Tr}\left\{H(|G\rangle\langle G|
- |\psi\rangle\langle\psi|) \right\}$ cannot be greater
than the product
\be\label{CS}\left\{\mbox{Tr}(H^2)\right\}^{1/2}
\left\{\mbox{Tr}(|G\rangle\langle G| -
|\psi\rangle\langle\psi|)^2\right\}^{1/2}.\ee The first
factor of (\ref{CS}) is equal to $\| {\bf E}\|$ and the
second factor is equal to the square root of $2(1-F^2)$. We
now find that \be \langle G|H|G\rangle &=&
\mbox{Tr}\left\{H(|G\rangle\langle G| -
|\psi\rangle\langle\psi|) \right\} +
\langle\psi|H|\psi\rangle\nonumber\\ &\leq&\sqrt{2} \| {\bf
E}\|(1-F^2)^{1/2} + \langle\psi|H|\psi\rangle\nonumber\\
&=& \sqrt{2} \| {\bf E}\|(1-F^2)^{1/2} + E_0.\ee Combining
this identity with (\ref{Ky}) proves the result. \finpr

An important implication of this result is the following.
Letting $\Delta E=E_1-E_0$ be the energy gap of $H$ between
the ground state energy and the first excited level, one
finds that $E_i\geq E_0 + \Delta E$ for every $i=1, \dots,
2^n-1$, and therefore \be \label{bound_gap} \frac{(r-1)\Delta E}{r}\leq
\frac{E_0+\dots + E_{r-1}}{r} - E_0 .\ee This shows that
\be\frac{r-1}{\sqrt{2}r} \frac{\Delta E}{\|{\bf E}\|}\leq
(1-|\langle G|\psi\rangle|^2)^{1/2}.\ee As $r\geq 2$ for
any $d<\eta(|G\rangle)$, this proves that, \emph{any ground
state of a $d$-body Hamiltonian $H$ with $d<
\eta(|G\rangle)$ can only be $\epsilon$-close to the graph
state $|G\rangle$ at the cost of $H$ having
an energy gap which is $\epsilon$-small  relative to the total energy in the system}.

In a first approximation, this result indicates it might be difficult to robustly create states
close to the graph state $|G\rangle$ by cooling a system
governed by a $d$-body Hamiltonian into its non-degenerate
ground state, as minor thermal fluctuations may easily
bring the system in an excited state (which is orthogonal
to the ground state), since the energy gap is necessarily
small. However, two important remarks regarding the precise interpretation of this result are in order.

First, eq. (\ref{bound_gap}) is stated in terms of the quantity $\Delta E_{\mbox{\scriptsize{rel}}}:=\Delta E/\|{\bf E}\|$, i.e., the energy gap relative to the total energy, rather than in terms of the absolute energy gap. This implies that, in physical systems where $\Delta E$ is held constant and where $\|{\bf E}\|$ is very large, one finds that $\Delta E_{\mbox{\scriptsize{rel}}}$ is arbitrarily small. This situation might e.g. occur if there exist a large number of energy levels in the system, each a constant distance $\Delta E$ apart. Hence, in such cases the bound (\ref{bound_gap}) does not seem to be very useful. The fact that $\Delta E_{\mbox{\scriptsize{rel}}}$  appears in (\ref{bound_gap}), and not the absolute gap $\Delta E$,  is due to the fact that the bound is totally general, in that it holds for \emph{all} Hamiltonians; in particular, it does not exclude situations where e.g. the ground state level is maximally degenerate.

Second, we also note that the fidelity might not be the best suited distance measure in certain applications. If one is for instance interested in creating 2D cluster states in order to build a one-way quantum computer, it is known that if the 2D cluster states are subject to local noise which is below a certain threshold, the resulting states still enable universal quantum computation---this follows from the fault tolerance of the one-way model, where computation is performed on encoded states \cite{Ra03'}. However, when the fidelity is used as a distance measure, the original 2D cluster states and the noisy ones can be very far apart (their fidelity can even be exponentially small). Thus, for such applications one should be cautious in using the bound (\ref{bound_gap}).

Finally, we note that (an analogue of) theorem \ref{thm_techn} can also be derived from
theorem 1 in Ref. \cite{Ha03}, where a general  bound on the
fidelity between an arbitrary state $|\phi\rangle$ and the
ground state $|\psi\rangle$ of a Hamiltonian $H$ is
obtained in terms of the spectrum of $H$.

In the next section we
consider Hamiltonians on $n'>n$ qubits, thus allowing for
ancilla particles, which are to be constructed in such a
way that the desired $n$-qubit graph states occur as states
on a subset of the initial system of $n'$ qubits.

\section{Ancilla particles and Gadget constructions}\label{sect_ancilla}

The goal of this section is to demonstrate how one can construct a 2-body Hamiltonian with approximately the same ground state as a given many-body Hamiltonian. The basic idea is from Refs. \cite{Ol05, Ke04}2 where they use perturbation theory to show that by adding ancilla qubits one can construct a 2-body Hamiltonian whose ground state is arbitrarily close to the ground state of any many-body Hamiltonian. By adding ancilla qubits we can avoid the problems discussed in the previous sections.

In our case we would like to find a 2-body Hamiltonian with a ground state close to the honeycomb lattice graph state. It is known that the honeycomb lattice graph state is a universal resource for measurement based quantum computation \cite{Va06}. Therefore, the 2-body Hamiltonian we construct in this section will have a universal resource for quantum computation as a ground state.

We start in section \ref{linear-cluster-ancilla} by examining one of the steps required in the construction. By using the linear cluster state as an example, we show how to find a 2-body Hamiltonian with approximately the same ground state as a given 3-body Hamiltonian. In section \ref{honeystate-ancilla} we expand on this and show how to find a 2-body Hamiltonian whose ground state is approximately the honeycomb lattice graph state. We conclude with section \ref{generic-ancilla} by discussing how any graph state can be the approximate ground state of a 2-body Hamiltonian.

\subsection{Linear cluster state}\label{linear-cluster-ancilla}

In order to create a 2-body Hamiltonian with the same ground state as a given 3-body Hamiltonian $H$,  we create a new perturbed Hamiltonian $\tilde{H} = K + V$ where $K$ is a Hamiltonian with large spectral gap and degenerate ground space associated with eigenvalue $0$. $V$ is a Hamiltonian with norm much smaller than the gap of $K$ that attempts to recreate the spectrum of $H$ in the gap of $K$ by raising the degeneracy.

Our starting Hamiltonian will be (\ref{H_C}) which has the linear cluster state on $n$ qubits as a ground state. We will only consider Hamiltonians with periodic boundary conditions. In this case, a linear cluster state in a circle. We construct a 2-body Hamiltonian with a ground state that is close to the ground state of (\ref{H_C}).  In order the apply the perturbation theory argument given in \cite{Ke04} we require our Hamiltonian to be in a certain form. We simply rewrite (\ref{H_C}) as
\begin{eqnarray}
H = \sum_{i=1}^n \frac{-n^9}{6}\left(H_{1}^{(i)} -6  B^{(i-1)}_z B^{(i)}_x B^{(i+1)}_z \right)
\end{eqnarray}
where
\begin{eqnarray}
B^{(i)}_{j} =  \left(\frac{2}{n^3}I + \frac{1}{n^3} \sigma^{(i)}_j \right)
\end{eqnarray}
and $H_{1}^{(i)}$ is a Hamiltonian that contains only 2-body, 1-body and identity terms, namely
\begin{eqnarray}
H_1^{(i)} &=& \frac{48}{n^9} I + \frac{24}{n^9} \left( \sigma_x^{(i)} + \sigma_z^{(i+1)}  + \sigma_z^{(i-1)}  \right)\\ \nonumber
&+& \frac{12}{n^9} \left( \sigma_x^{(i)}\sigma_z^{(i+1)} + \sigma_x^{(i)}\sigma_z^{(i-1)} +
\sigma_z^{(i-1)}\sigma_z^{(i+1)} \right).
\end{eqnarray}
Let $i_1, i_2$ and, $i_3$ be the ancilla qubits for qubit $i$. For a sufficiently small $\delta$ we can show that the 2-body perturbed Hamiltonian $\tilde{H} = -\frac{n^9}{6} ( K + V )$ has approximately the same ground state as (\ref{H_C}), where
\begin{eqnarray}\label{Hamiltonian_K}
K &=& -\frac{\delta^{-3}}{4} \sum_{i=1}^n \left(\sigma_z^{(i_1)}\sigma_z^{(i_2)}  + \sigma_z^{(i_1)}\sigma_z^{(i_3)} \right) \nonumber \\
&-& \frac{\delta^{-3}}{4} \sum_{i=1}^n \left( \sigma_z^{(i_2)}\sigma_z^{(i_3)} - 3I \right)
\end{eqnarray}
and
\begin{eqnarray}
\nonumber V \!\!&=&\!\! \sum_{i=1}^n \!H_1^{(i)} + \delta^{-1} \sum_{i=1}^n \left( B^{(i-1)2}_x \!+\! B^{(i)2}_z \!+\! B^{(i+1)2}_z \right) \\
&-&  \delta^{-2} \sum_{i=1}^n \left( B^{(i-1)}_z  \sigma_x^{i_1}+ B^{(i)}_x \sigma_x^{i_2} + B^{(i+1)}_z \sigma_x^{i_3} \right)\!.
\end{eqnarray}
Notice that the Hamiltonian $K$ has eigenvalues $0$ and $\delta^{-3}$. We use $V$ to manipulate this gap in order to approximate (\ref{H_C}).

By substituting in the appropriate values for the $B^{(i)}_j$ in $V$ we get our final 2-body Hamiltonian, $\tilde{H} = -\frac{n^9}{6} ( K + V)$ where $K$ is as above and
\begin{eqnarray}\label{coeff_space}
\nonumber
V \!\!&=&\!\! \sum_{i=1}^n \!H_1^{(i)} + \frac{\delta^{-1}}{n^6} \sum_{i=1}^n \left( 15 I \!+ \!4 \sigma_z^{(i-1)} \!+\! 4 \sigma_x^{(i)} \!+\! 4 \sigma_z^{(i+1)} \right) \\ \nonumber
&-& \frac{\delta^{-2}}{n^3} \sum_{i=1}^n \left( 2 \sigma_x^{(i_1)} + 2 \sigma_x^{(i_2)}  + 2 \sigma_x^{(i_3)} \right) \\
&-& \frac{\delta^{-2}}{n^3} \sum_{i=1}^n \left( \sigma_x^{(i-1)} \sigma_x^{i_1} + \sigma^{(i)}_z  \sigma_x^{i_2}+ \sigma^{(i+1)}_z \sigma_x^{i_3} \right).
\end{eqnarray}

\begin{figure}
\begin{center}
\includegraphics[width=85mm]{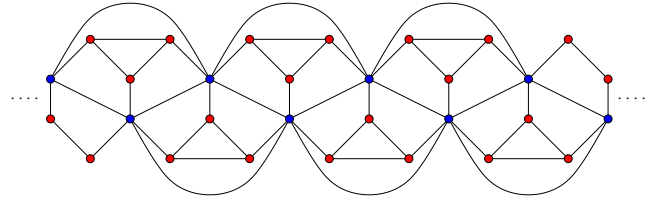}
\caption{An interaction diagram of a 2-body Hamiltonian with the linear cluster state as the ground state. The original qubits are shown in blue and the red qubits are the added ancilla qubits.}
\label{fig:linear_cluster}
\end{center}
\end{figure}

Figure \ref{fig:linear_cluster} shows the interaction diagram of the linear cluster state which is the
approximate ground state of a 2-body Hamiltonian. The original qubits are shown in blue and
the red qubits are the added ancilla qubits. This diagram is significantly less complex than the interaction diagram of a 2-body Hamiltonian constructed from the 4-body honeycomb lattice, which we consider below.

Although this section has shown us a method for creating a 2-body Hamiltonian with approximately the same ground state as a 3-body Hamiltonian, it has also demonstrated a major disadvantage with this method. In order for the perturbation theory to apply we need to choose a small $\delta$. The smaller $\delta$ is the greater the difference between the coefficients of $K$ and the parts of $V$ becomes.
As this gap becomes bigger it becomes more difficult to create such a Hamiltonian in the laboratory as it requires a high degree of precision over several orders of magnitude. Added to this problem is the fact that the coefficients depend on the number of qubits in the original lattice, see eqn. (\ref{coeff_space}) for an example. As the number of qubits increases the difference between the coefficients grows even larger.

\subsection{Honeycomb lattice graph state}\label{honeystate-ancilla}

Next, we use the gadgets from above to show that the honeycomb lattice graph state can occur as the approximate ground state of a 2-body Hamiltonian. We show this by using a two step process. We start with a 4-body Hamiltonian with the honeycomb lattice graph state as a ground state. Then, using results from \cite{Ol05} we create a 3-body Hamiltonian with approximately the same ground state. We then use the procedure described above to create a 2-body Hamiltonian with approximately the same ground state as our 3-body Hamiltonian.

Our starting Hamiltonian is given by
\begin{eqnarray}
H = - \sum_{i} \sigma_x^{(i)} \sigma_z^{(i_a)} \sigma_z^{(i_b)} \sigma_z^{(i_c)},
\end{eqnarray}
where $i_a, i_b$, and $i_c$ are the neighbouring qubits to qubit $i$ in the hexagonal lattice graph and each term is a member of the generator for the stabilizer of the graph state.
Is it clear that the ground state of this 4-body Hamiltonian is the honeycomb lattice state. We note that the hexagonal lattice graph state is a universal resource for measurement based quantum computation.

The first step requires us to add one ancilla qubit for each qubit in our lattice. We split each 4-body term into two 2-body terms and then couple each of these to the ancilla qubit. When done we have two 3-body terms for each 4-body term in our original Hamiltonian. This process can be used to create a Hamiltonian with a ground state that approximates the ground state of a d-body Hamiltonian but has only $\lceil d/2 \rceil +1$-body interactions.

As shown in the linear cluster state example above, the second step requires we add three ancilla qubits for each 3-body term in the Hamiltonian we are trying to approximate. This means that the final ground state of the 2-body Hamiltonian will have 7 ancilla qubits for each qubit in the original ground state.

After applying the two steps above the resulting Hamiltonian is
\begin{eqnarray}
\tilde{H} = - a \sum_{i} \left( A_i + b B_i + c C_i + d D_i + e I\right),
\end{eqnarray}
where $A_i$, $B_i$, $C_i$, and $D_i$ are defined below and $a, b, c, d$, and $e$ are constants depending on $n$ and the spectral gap of the Hamiltonian $H$.

In the 7 added ancilla qubits there are two sets of three qubits which interact. The interaction diagram for each of the two sets forms a triangle. The first part of our Hamiltonian, $A_i$, is the part that describes the interactions between these six ancilla qubits. Let $i_1$, $i_2$, $i_3$, $i_4$, $i_5$, and $i_6$ be six ancilla qubits for qubit $i$.
Then,
\begin{eqnarray}
A_i &=& \sigma_z^{(i_1)}\sigma_z^{(i_2)} +  \sigma_z^{(i_1)}\sigma_z^{(i_3)} +  \sigma_z^{(i_2)}\sigma_z^{(i_3)} \nonumber \\ &+& \sigma_z^{(i_4)}\sigma_z^{(i_5)} +  \sigma_z^{(i_4)}\sigma_z^{(i_6)} +  \sigma_z^{(i_5)}\sigma_z^{(i_6)}.
\end{eqnarray}

The second part of our 2-body Hamiltonian functions in much the same way as the first part. This part of the Hamiltonian is a result of the first step when we created a 3-body Hamiltonian. We partitioned each original four part stabilizer term into two parts, each with two operators. We then coupled
each pair with a common ancilla essentially forming
triangles which share a common vertex, the seventh ancilla qubit. We have,
\begin{eqnarray}
B_i &=& \sigma_x^{(i)}\sigma_z^{(i_a)} +  \sigma_x^{(i)}\sigma_x^{(i_7)} +  \sigma_z^{(i_a)}\sigma_x^{(i_7)} \nonumber \\ &+& \sigma_z^{(i_a)}\sigma_z^{(i_b)} +  \sigma_z^{(i_b)}\sigma_x^{(i_7)} +  \sigma_z^{(i_c)}\sigma_x^{(i_7)}.
\end{eqnarray}

The correlations described by $C_i$ are those correlations which connect the triangles created in $A_i$ and $B_i$. Each vertex of a triangle from $A_i$ is coupled to a vertex from a triangle in $B_i$. $C_i$ can be written as
\begin{eqnarray}
C_i &=& \sigma_x^{(i)}\sigma_x^{(i_1)} +  \sigma_z^{(i_a)}\sigma_x^{(i_2)} +  \sigma_x^{(i_3)}\sigma_x^{(i_7)} \nonumber \\ &+& \sigma_z^{(i_b)}\sigma_x^{(i_4)} +  \sigma_z^{(i_c)}\sigma_x^{(i_5)} +  \sigma_x^{(i_6)}\sigma_x^{(i_7)}.
\end{eqnarray}

The final part of our 2-body Hamiltonian describes the remaining operators acting on single qubits,
\begin{eqnarray}
D_i &=& \sigma_x^{(i)} + \sigma_z^{(i_a)} + \sigma_z^{(i_b)} + \sigma_z^{(i_c)} + 2 \sigma_z^{(i_7)} \nonumber \\ &+&
 d_1 \sum_{j=1}^6 \sigma_x^{(i_j)}  + d_2 \ketbra{1}{1}^{(i_7)}.
 \end{eqnarray}
Again, the constants $d_1$ and $d_2$ depend on the number of qubits and the spectral gap of the Hamiltonian $H$.

\subsection{Generic graph states}\label{generic-ancilla}

Given any graph state $|G\ket$ it is always possible to create a 2-body Hamiltonian that has $|G\ket$ (along with some ancilla qubits) as the approximate ground state. If $|G\ket$ is a graph state on $n$ qubits then we know from section IV that there exists a set $\{g_1, g_2, \dots , g_n \}$ of generators for the
stabilizer of $|G\ket$ such that each generator has weight at most $\eta(|G\ket)$. We can construct the $\eta(|G\ket)$-body Hamiltonian
\begin{eqnarray}
H = - \sum_{i=1}^n g_i
\label{make2body}
\end{eqnarray}
which has $|G\ket$ as the ground state. We then use the steps described above to reduce this Hamiltonian to one which has only 2-body interactions. For every $k$-body interaction in the Hamiltonian (\ref{make2body}) we require an additional $O(k)$ ancilla qubits for our 2-body Hamiltonian. In the worst case we would need to add $O(\eta(|G\ket) n)$ ancilla qubits.

In this section we have shown that by adding ancilla qubits we can overcome the problems we addressed in the previous sections. We have shown that one can construct a 2-body Hamiltonian
whose ground state is close to a universal resource for measurement based quantum computation. Unfortunately, the Hamiltonians we have constructed are only of theoretical interest. Due to the high degree of control and precision that is required to create these Hamiltonians they are of little value for practical applications. Finally, we note that a bound similar to (\ref{bound_gap}) can also be obtained for the gadget construction. Note that in this situation, the total energy $\|{\bf E} \|$ is typically large \footnote{This is e.g. reflected in equation (\ref{Hamiltonian_K}), where the Hamiltonian $K$ has a large prefactor $d^{-3}$ (since $\delta$ is small) and therefore has a large norm. }, such that the relative gap $\Delta E/ \|{\bf E} \|$ is small.

\section{Conclusion}\label{sect_concl}

In this paper we have settled the issue whether graph
states can occur as ground states of two-body Hamiltonians.
More generally, we have shown that the quantity
$\eta(|G\rangle)$, defined as the minimal $d$ such that
${\mathcal S}_d = {\mathcal S}$, is of central interest in
the present context. It determines the minimum number of
interactions in the sense that $\eta(|G\rangle)$-body
interaction are required to obtain $|G\rangle$ as exact,
non-degenerate ground state. In addition, we found that
$\eta(|G\rangle) \geq 3$ for all graph states, which
implies that any $n$-qubit graph state $|G\rangle$ cannot
be the exact non-degenerate ground state of a two-body
Hamiltonian acing on $n$ qubits. We have also related the
accuracy $\epsilon$ of approximating the graph state
$|G\rangle$ using a Hamiltonian with $d'$-body interactions
and $d' < \eta(|G\rangle)$, to the energy gap of the
Hamiltonian relative to the total energy, which turns out to be proportional to
$\epsilon$. When allowing the usage of ancilla particles
that act as mediating particles to generate an effective
many-body Hamiltonian on a subsystem, we have shown that
the gadget construction introduced in Refs. \cite{Ke04, Ol05} can be used
to obtain the $n$-qubit graph state $|G\rangle$ as an
non-degenerate (quasi) ground state of a two-body
hamiltonian acting on $n'> n$ qubits. However, an
incredible high accuracy in the control of the parameters
of the interaction hamiltonian is required.
We also remark that our result do not directly apply to the generation of graph states in an {\em encoded} form. On the one hand, an (exponential) small fidelity of the physical state might still be acceptable to obtain high fidelity with respect to the encoded (logical) graph states when using redundant encodings corresponding to quantum error correcting codes. On the other hand, as demonstrated in Ref. \cite{Ba06}, there exist  (approximate) ground states of two--body hamiltonians which are arbitrary close to encoded graph states with respect to a certain encoding. The energy gap of the corresponding Hamiltonian is constant, independent of the system size, and the encoded graph states constitute a universal resources for measurement based quantum computation using only single qubit measurements. The usage of encoded graph states for measurement based quantum computation is subject of ongoing research \cite{Gr06,DuVa06}.

We finally remark that the quantity $\eta(|G\rangle)$
serves as a natural complexity measure of graph states, as
it assesses how difficult it is to exactly prepare a state by
cooling a system into its ground state. The present results
show that graph states typically exhibit a large complexity
in this sense, whereas they have small computational
complexity, since all graph states can be prepared with a
poly-sized quantum circuit.

\section*{Acknowledgements}
We thank J. Kempe for useful discussions. This work was supported by the Austrian Science Foundation (FWF),
the European Union (QICS,OLAQUI,SCALA), and the Austrian Academy of Sciences
(\"OAW) through project APART (W.D.).


\begin{thebibliography}{99}



\bibitem{He06}
M.\ Hein {\em et al.}, Proceedings of the International
School of Physics ``Enrico Fermi'' on ``Quantum Computers,
Algorithms and Chaos'',  Varenna, Italy, July, 2005,
quant-ph/0602096.


\bibitem{Br01}
H. J. Briegel and R. Raussendorf, Phys. Rev. Lett. {\bf
86}, 910 (2001).

\bibitem{Ra01}
R. Raussendorf and H. J. Briegel, Phys. Rev. Lett. {\bf
86}, 5188 (2001); Quantum Inf. Comp. {\bf 2}(2), 443
(2002).


\bibitem{Ra03}
R Raussendorf, D. E. Browne, and H. J. Briegel, Phys. Rev.
A {\bf 68}, 022312 (2003).


\bibitem{Go97}
D. Gottesman, PhD thesis, Caltech, 1997.

\bibitem{Ca96}
A. R. Calderbank and P. W. Shor, Phys. Rev. A {\bf 54}, 1098 (1996).

\bibitem{St96}
A. M. Steane, Phys. Rev. Lett. {\bf 77}, 793 (1996).

\bibitem{Hi99}
M. Hillery, V. Buzek, and A. Berthiaume, Phys. Rev. A {\bf 59}, 1829 (1999).

\bibitem{Ch04}
Kai Chen and Hoi-Kwong Lo, E-print: quant-ph/0404133.

\bibitem{Du05}
W. D\"ur, J. Calsamiglia and H.-J. Briegel, Phys. Rev. A {\bf 71}, 042336 (2005).


\bibitem{Ni05}
M. A. Nielsen, quant-ph/0504097.

\bibitem{Ba06}
S. D. Bartlett and T. Rudolph, Phys. Rev. A. {\bf 74}, 040302(R) (2006).

\bibitem{Va06}
M. Van den Nest, A. Miyake, W. D\"ur and H. J. Briegel,
Phys. Rev. Lett. {\bf 97}, 150504 (2006).

\bibitem{Mh04}
M. Mhalla, S. Perdrix, quant-ph/0412071.

\bibitem{Va04}
M. Van den Nest, J. Dehaene, B. De Moor, Phys. Rev. A {\bf 69}, 022316 (2004), quant-ph/0308151.

\bibitem{foota}
A set of stabilizer elements is called \emph{independent}
if no nontrivial product of operators in this set yields
the identity.

\bibitem{foot}
In other words, ${\cal S}_d$ is the set of all elements $g$
in ${\cal S}$ which can be written as a product $g=g_1\dots
g_k$ for some $g_1, \dots, g_k\in {\cal S}$ and for some
$k$, such that wt$(g_i)\leq d$ for all $i=1, \dots k$.


\bibitem{Ha03}
H. L. Haselgrove, M. A. Nielsen, T. J. Osborne, Phys. Rev.
A {\bf  69} (3), 032303 (2004), quant-ph/0308083.

\bibitem{Ke04}
Julia Kempe, Alexi Kitaev, Oded Regev, SIAM Journal of Computing, Vol. {\bf 35}(5), p. 1070-1097 (2006)

\bibitem{Ol05}
Roberto Oliveira, Barbara M. Terhal, quant-ph/0504050

\bibitem{Gr06}
D. Gross and J. Eisert, quant-ph/0609149.

\bibitem{DuVa06}
W. D\"ur, M. Van den Nest, A. Miyake, and H.-J. Briegel, in preparation.

\bibitem{Ra03'}
R. Raussendorf, PhD thesis, LMU Munich (2003).

\end{thebibliography}
\end{document}